\documentclass{PoS}

\usepackage{ifthen} 
\newboolean{pdflatex}
\setboolean{pdflatex}{true} 

\newboolean{articletitles}
\setboolean{articletitles}{true} 

\newboolean{uprightparticles}
\setboolean{uprightparticles}{false} 

\newboolean{inbibliography}
\setboolean{inbibliography}{false} 

\usepackage{xspace} 
\usepackage{upgreek}

\def\MagUp {\mbox{\em Mag\kern -0.05em Up}\xspace}

\ifthenelse{\boolean{uprightparticles}}%
{

 \def\PDelta      {\ensuremath{\Delta}\xspace}                 
 \def\PXi      {\ensuremath{\Xi}\xspace}                 
 \def\PLambda      {\ensuremath{\Lambda}\xspace}                 
 \def\PSigma      {\ensuremath{\Sigma}\xspace}                 
 \def\POmega      {\ensuremath{\Omega}\xspace}                 
 \def\PUpsilon      {\ensuremath{\Upsilon}\xspace}                 
 
 \def\PB      {\ensuremath{\mathrm{B}}\xspace}                 
                  
 \def\PD      {\ensuremath{\mathrm{D}}\xspace}

 \def\PK      {\ensuremath{\mathrm{K}}\xspace}

 \def\Pb      {\ensuremath{\mathrm{b}}\xspace}

 \def\Pi      {\ensuremath{\mathrm{i}}\xspace}

 \def\Ps      {\ensuremath{\mathrm{s}}\xspace}

}
{

 \mathchardef\PDelta="7101
 \mathchardef\PXi="7104
 \mathchardef\PLambda="7103
 \mathchardef\PSigma="7106
 \mathchardef\POmega="710A
 \mathchardef\PUpsilon="7107
                  
 \def\PB      {\ensuremath{B}\xspace}                 
                  
 \def\PD      {\ensuremath{D}\xspace}

 \def\PK      {\ensuremath{K}\xspace}

 \def\Pb      {\ensuremath{b}\xspace}

 \def\Pi      {\ensuremath{i}\xspace}

 \def\Ps      {\ensuremath{s}\xspace}

}

\makeatletter
\ifcase \@ptsize \relax
  \newcommand{\miniscule}{\@setfontsize\miniscule{4}{5}}
\or
  \newcommand{\miniscule}{\@setfontsize\miniscule{5}{6}}
\or
  \newcommand{\miniscule}{\@setfontsize\miniscule{5}{6}}
\fi
\makeatother

\DeclareRobustCommand{\optbar}[1]{\shortstack{{\miniscule (\rule[.5ex]{1.25em}{.18mm})}
  \\ [-.7ex] $#1$}}

\def\squark    {{\ensuremath{\Ps}}\xspace}

\def\bquark    {{\ensuremath{\Pb}}\xspace}

  \def\Kbar    {{\kern 0.2em\overline{\kern -0.2em \PK}{}}\xspace}

\def\KorKbar    {\kern 0.18em\optbar{\kern -0.18em K}{}\xspace}

  \def\Dbar    {{\kern 0.2em\overline{\kern -0.2em \PD}{}}\xspace}
\def\D       {{\ensuremath{\PD}}\xspace}

\def\DorDbar    {\kern 0.18em\optbar{\kern -0.18em D}{}\xspace}
\def\Dz      {{\ensuremath{\D^0}}\xspace}

\def\Ds      {{\ensuremath{\D^+_\squark}}\xspace}

\def\B       {{\ensuremath{\PB}}\xspace}
\def\Bbar    {{\ensuremath{\kern 0.18em\overline{\kern -0.18em \PB}{}}}\xspace}

\def\BorBbar    {\kern 0.18em\optbar{\kern -0.18em B}{}\xspace}
\def\Bz      {{\ensuremath{\B^0}}\xspace}

\def\Bu      {{\ensuremath{\B^+}}\xspace}

\def\Bp      {{\ensuremath{\Bu}}\xspace}

\def\Bs      {{\ensuremath{\B^0_\squark}}\xspace}

  \def\Y#1S{\ensuremath{\PUpsilon{(#1S)}}\xspace}

\def\Lz          {{\ensuremath{\PLambda}}\xspace}
\def\Lbar        {{\ensuremath{\kern 0.1em\overline{\kern -0.1em\PLambda}}}\xspace}
\def\LorLbar    {\kern 0.18em\optbar{\kern -0.18em \PLambda}{}\xspace}

\def\Lb      {{\ensuremath{\Lz^0_\bquark}}\xspace}

\def\to                 {\ensuremath{\rightarrow}\xspace}

\def\AT#1     {\ensuremath{A_{\mathrm{T}}^{#1}}\xspace}           

\def\C#1      {\ensuremath{\mathcal{C}_{#1}}\xspace}                       
\def\Cp#1     {\ensuremath{\mathcal{C}_{#1}^{'}}\xspace}                    
\def\Ceff#1   {\ensuremath{\mathcal{C}_{#1}^{\mathrm{(eff)}}}\xspace}        
\def\Cpeff#1  {\ensuremath{\mathcal{C}_{#1}^{'\mathrm{(eff)}}}\xspace}       
\def\Ope#1    {\ensuremath{\mathcal{O}_{#1}}\xspace}                       
\def\Opep#1   {\ensuremath{\mathcal{O}_{#1}^{'}}\xspace}                    

\newcommand{\tev}{\ifthenelse{\boolean{inbibliography}}{\ensuremath{~T\kern -0.05em eV}\xspace}{\ensuremath{\mathrm{\,Te\kern -0.1em V}}}\xspace}
\newcommand{\gev}{\ensuremath{\mathrm{\,Ge\kern -0.1em V}}\xspace}
\newcommand{\mev}{\ensuremath{\mathrm{\,Me\kern -0.1em V}}\xspace}
\newcommand{\kev}{\ensuremath{\mathrm{\,ke\kern -0.1em V}}\xspace}
\newcommand{\ev}{\ensuremath{\mathrm{\,e\kern -0.1em V}}\xspace}
\newcommand{\gevc}{\ensuremath{{\mathrm{\,Ge\kern -0.1em V\!/}c}}\xspace}
\newcommand{\mevc}{\ensuremath{{\mathrm{\,Me\kern -0.1em V\!/}c}}\xspace}
\newcommand{\gevcc}{\ensuremath{{\mathrm{\,Ge\kern -0.1em V\!/}c^2}}\xspace}
\newcommand{\gevgevcccc}{\ensuremath{{\mathrm{\,Ge\kern -0.1em V^2\!/}c^4}}\xspace}
\newcommand{\mevcc}{\ensuremath{{\mathrm{\,Me\kern -0.1em V\!/}c^2}}\xspace}

\def\invfb   {\ensuremath{\mbox{\,fb}^{-1}}\xspace}

\def\gsim{{~\raise.15em\hbox{$>$}\kern-.85em
          \lower.35em\hbox{$\sim$}~}\xspace}
\def\lsim{{~\raise.15em\hbox{$<$}\kern-.85em
          \lower.35em\hbox{$\sim$}~}\xspace}

\def\tell1  {TELL1\xspace}
\def\ukl1   {UKL1\xspace}

\def\Bem{ $\B^0_{(s)}\rightarrow e^{\pm}\mu^{\mp}$\xspace}
\def\Bdem{ $\B^0\rightarrow e^{\pm}\mu^{\mp}$\xspace}
\def\Bsem{ $\B^0_{s}\rightarrow e^{\pm}\mu^{\mp}$\xspace}
\def\Dem{ $\Dz\rightarrow e^{\pm}\mu^{\mp}$\xspace}

\def\tmmm{ $\tau^{\pm}\to \mu^{\pm}\mu^{\mp}\mu^{\pm}$\xspace}

\usepackage[left]{lineno}

\usepackage{amsmath}
\DeclareMathAlphabet{\mathcal}{OMS}{cmsy}{m}{n}

\title{Lepton flavour and lepton number violation searches at the LHCb experiment}

\ShortTitle{Lepton flavour and lepton number violation searches at the LHCb experiment}

\author{\speaker{Luca Pescatore} on behalf of the LHCb collaboration
\\
        \'Ecole Polytechnique F\'ed\'erale de Lausanne\\
        E-mail: \email{luca.pescatore@cern.ch}}

\abstract{Recent hints for lepton-universality violation in $b\to c\ell\nu$ and $b\to s\ell\ell$ transitions could imply the existence of lepton-flavour violating $b$ decays. 
The LHCb experiment is well suited for the search for these decays due to its large acceptance and trigger efficiency, as well as its excellent invariant 
mass resolution and particle identification capabilities. Recent results on searches for lepton-flavour violating decays from the LHCb experiment are presented.}

\FullConference{XXXIX International Conference on High Energy Physics\\
		3-11 July 2018\\
		Seoul, South Korea}

\begin{document}

The conservation of lepton flavour  is well established by the non-observation of decays such as $\mu\to e\gamma$.
However, it is not supported by strong theoretical reasons and it is already violated by the observation of neutrino oscillations.
At the LHCb experiment, several searches for Lepton Flavour Violating (LFV) decays are performed including $B$-meson decays,
but also $\Lambda_b$, charmed hadrons and lepton decays. These measurements are particularly interesting because anomalies in 
Lepton Flavour Universality (LFU) tests of the Standard Model (SM) were observed~\cite{Aaij:2014ora,Aaij:2017vbb,Aaij:2015qmj}.
In several models, such as those involving lepto-quarks, LFU can be linked to LFV~\cite{Hiller:2016kry} and the branching ratios 
of flavour violating $b$ decays are predicted in the range \mbox{$10^{-8}-10^{-10}$}.

\section{\Dem}

The \Dem decay is predicted in extensions of the SM, for example in R-parity violating SUSY models, with a branching ratio that could be as high of $10^{-6}$~\cite{Burdman:2001tf}.
At LHCb this decay is searched using the full dataset from Run I, corresponding to 3~\invfb of integrated luminosity~\cite{Aaij:2015qmj}.
In order to reduce the combinatorics \Dz are tagged via the $D^{*}\to\Dz\pi$ decay.
The main challenge of this analysis is to distinguish signal and background from $\Dz \to\pi^{\pm}\pi^{\mp}$ decays.
Most of such decays are rejected using particle identification requirements but an irreducible component, mainly due to the pions decaying in flight into muons, remains.
As a further handle agains this background the fit to extract the yield is performed in two dimensions: the invariant mass of the $e\mu$ system, $m(e\mu)$,
and the difference between the $D^{*}$ and  $\Dz$ masses. The result of the fit is shown in Fig.~\ref{fig:Demu}.
No significant signal excess is observed and a limit is set on the branching ratio using the CLs method at $\mathcal{B}( \Dz\rightarrow e^{\pm}\mu^{\mp}) < 1.3\times10^{-8}$, at 90\% 
confidence level, using the abundant $\Dz\to K^{\pm}\pi^{\mp}$ decay for normalisation.
This is the best limit to date on this decay. 

\begin{figure}
\label{fig:Demu}
\centering
\includegraphics[width=0.49\textwidth]{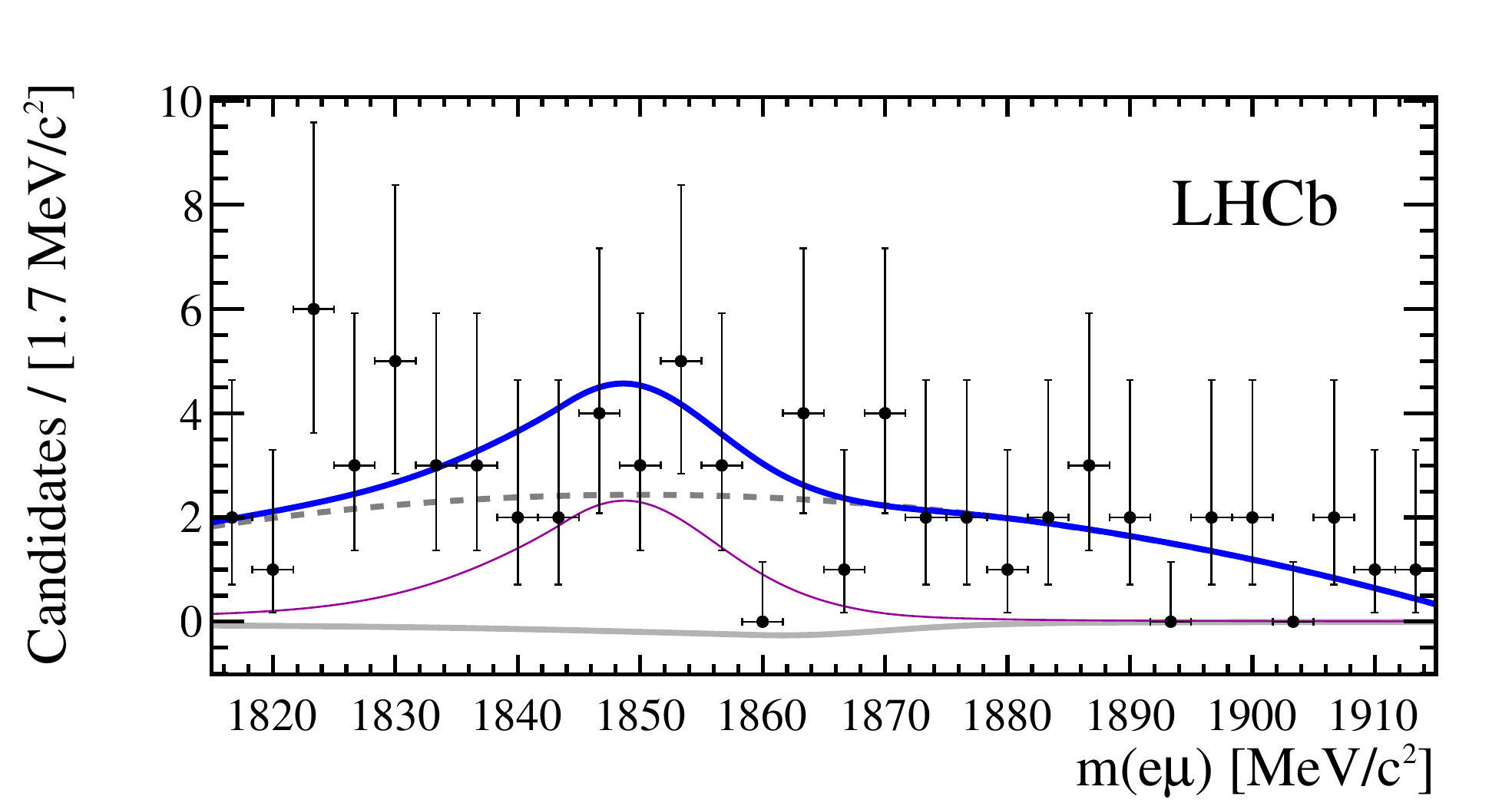}
\includegraphics[width=0.49\textwidth]{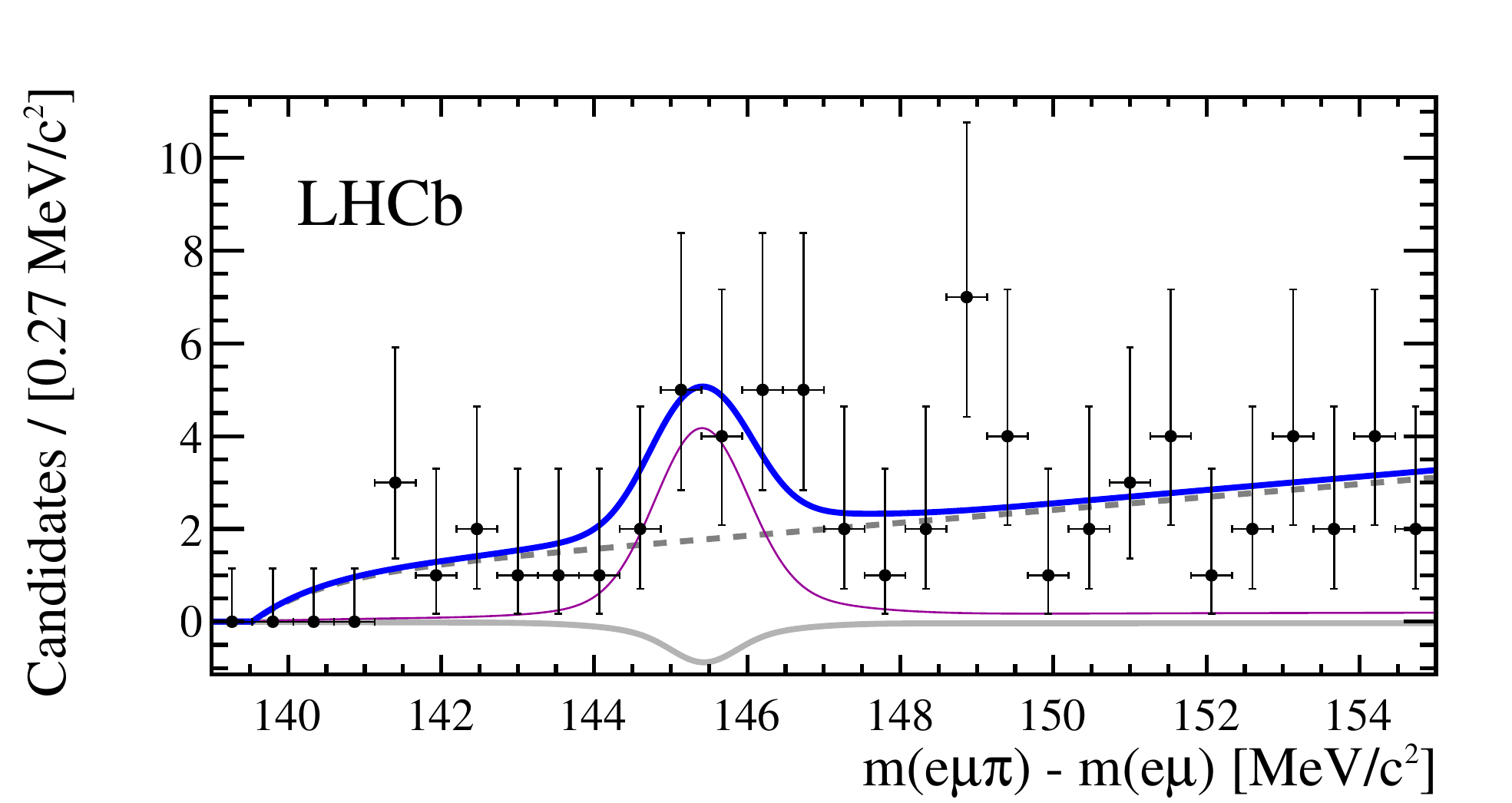}
\caption{ Invariant mass distribution of the reconstructed \Dem candidates (right).
The difference between the reconstructed $D^{*}$ and \Dz masses (left).}
\label{fig:Demu}
\end{figure}

\section{\Bem}

The \Bsem and \Bdem decays are also searched at LHCb using 3~\invfb of data collected in \mbox{Run I~\cite{Aaij:2017cza}}.
This analysis includes a special treatment of bremsstrahlung radiation emitted by the electron in the decay. Due to the bending of charged tracks, photons 
emitted before the calorimeter are absorbed by a different cell than the electron affecting the measurement of its momentum. 
A photon-recovery algorithm looks for clusters in the calorimeter which can be compatible with originating from a nearby electron track.
The signal candidate invariant mass distribution shapes depend on the number of recovered photons. Furthermore, the background levels also differ,
in fact, if a particle emitted photons, it is likely to be a real electron.
For this reasons the analysis is done separately on two bremsstrahlung categories which are fitted simultaneously.
The BDT is trained on simulated signal candidates and constructed such that its response for the signal decays is uniformly distributed between 0 and 1. 
The realistic BDT response to the signal is measured on data using $\Bz\to K^{\pm}\pi^{\mp}$ candidates as a signal proxy as shown in Fig.~\ref{fig:Bsemu}.
No cut is applied on the classifier output but the analysis is done in bins of it, which are fitted simultaneously.
This method allows to retain a better sensitivity and furthermore it reduces the systematic uncertainty due to the estimation of the
efficiency as this is calibrated on data rather then obtained from simulation.
Figure~\ref{fig:Bsemu} shows the fitted invariant mass distribution of the $e\mu$ system integrated over all considered categories.
No excess is observed and limits are set at $\mathcal{B}( \Bs\rightarrow e^{\pm}\mu^{\mp} ) < 5.4 ( 6.3 ) \times 10 ^{-9}$ 
and $\mathcal{B}( \Bz \rightarrow e^{\pm}\mu^{\mp} ) <1.0 (1.3) \times 10^{-9}$ at 90\% (95\%) confidence level.
These represent the best limits to date on these channels and the first limit for the $\Bs$ channel.

\begin{figure}
\label{fig:Bsemu}
\centering
\includegraphics[width=0.48\textwidth]{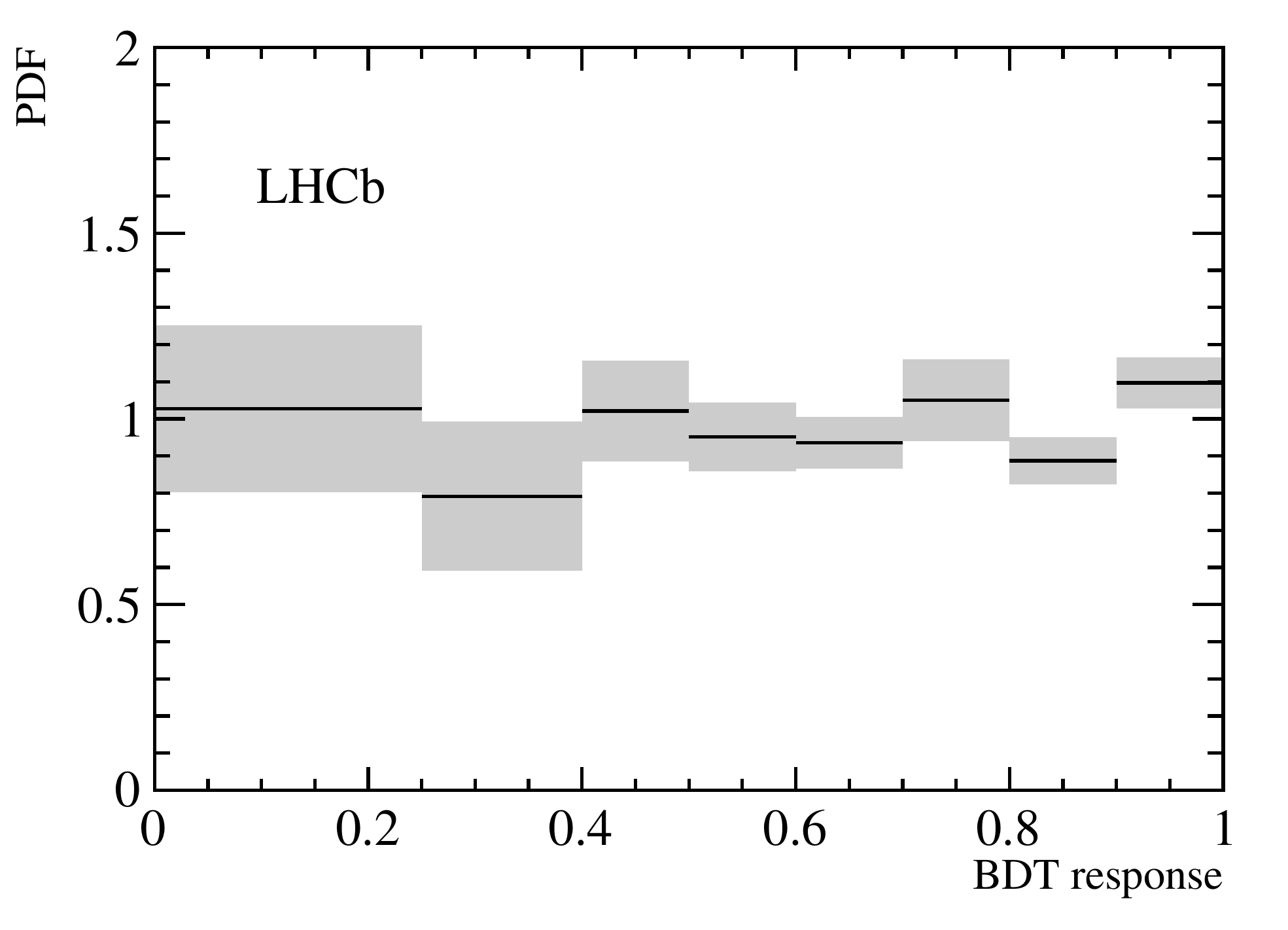}
\includegraphics[width=0.48\textwidth]{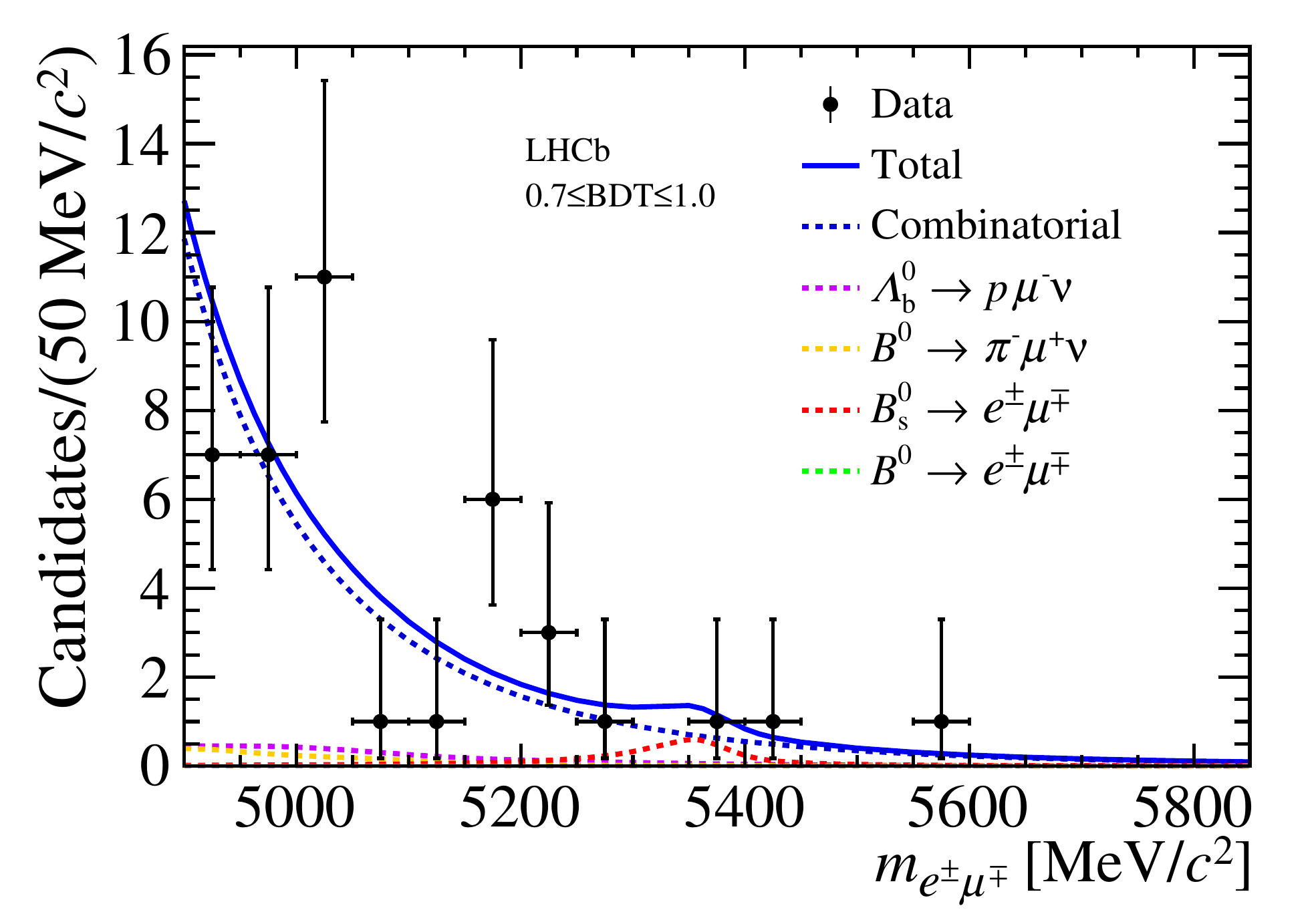}
\caption{ Fraction of $\Bz\to K^{\pm}\pi^{\mp}$ decays as a function of the classifier output.  (left)
Invariant mass distribution of the $e\mu$ system for \Bem candidates integrated over all categories considered. (right) }
\label{fig:Bsemu}
\end{figure}


\section{\tmmm}

The \tmmm decay was analysed at LHCb using 3~\invfb of data collected in Run I~\cite{Aaij:2014azz}.
Events are studied in a binned three dimensional space where the dimensions are: the mass of the $\tau$ candidate
and the output of two multivariate classifiers: one aimed at rejecting combinatorial background and based on topological information such as vertex quality 
and displacement and the second aimed at rejecting mis-identified decays and based on particle identification information.
The $\Ds\to(\phi\to\mu^{\pm}\mu^{\mp})\pi^{+}$ decay is used for normalisation and its invariant mass is shown in Fig.~\ref{fig:tau3mu} 
together with the invariant mass of the $\tau$ candidates. No signal excess is observed and a limit is set at 
$\mathcal{B}(\tau^{\pm}\to \mu^{\pm}\mu^{\mp}\mu^{\pm}) < 4 \times 10^{-8}$ at 90\% confidence level, which is competitive with the best limit set by Belle~\cite{Hayasaka:2010np}.
Previously, LHCb had set limits also on lepton number and baryon number violating decays of $\tau$~\cite{Aaij:2013fia}.

\begin{figure}
\label{fig:tau3mu}
\centering
\includegraphics[width=0.48\textwidth]{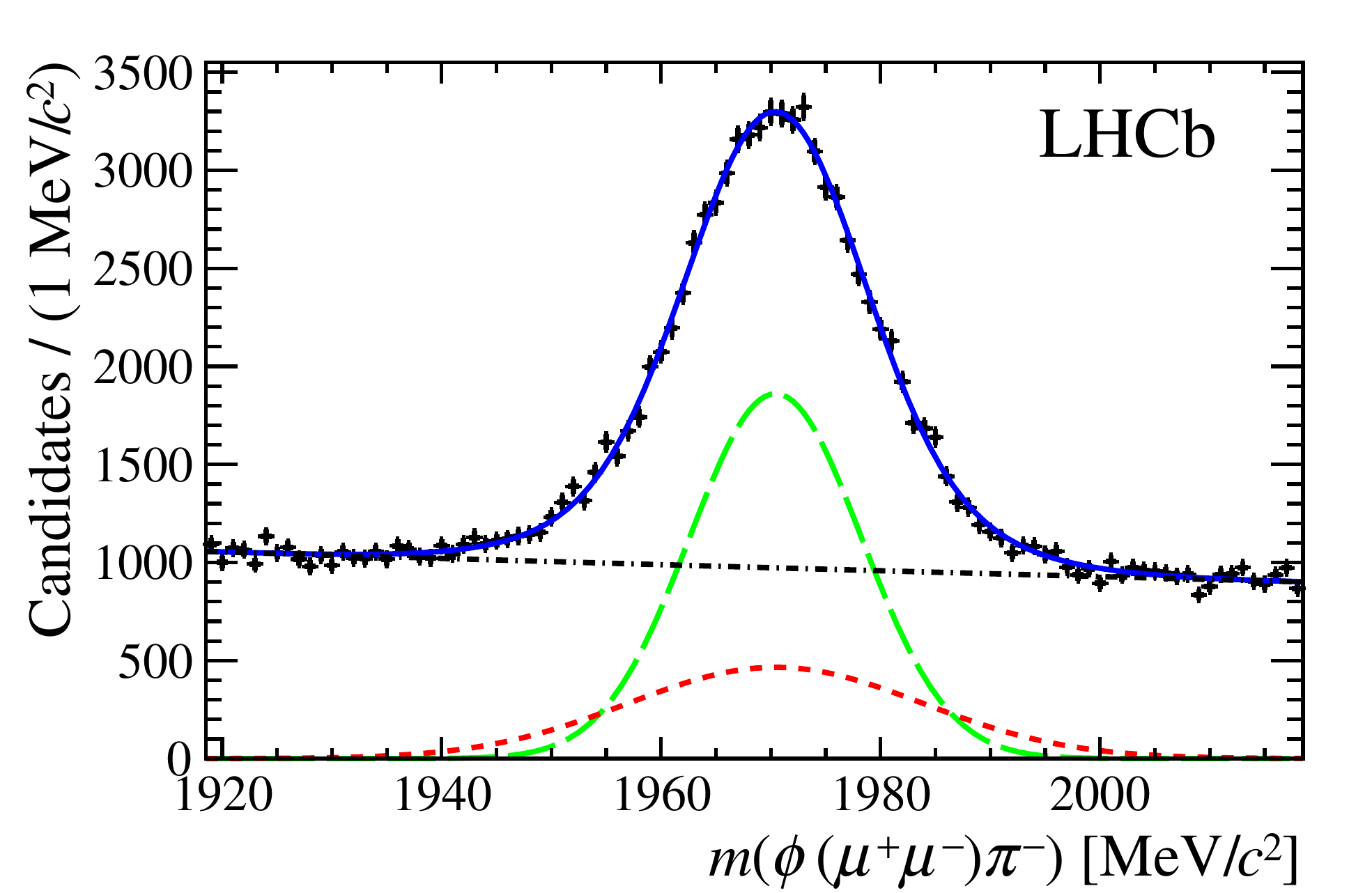}
\includegraphics[width=0.48\textwidth]{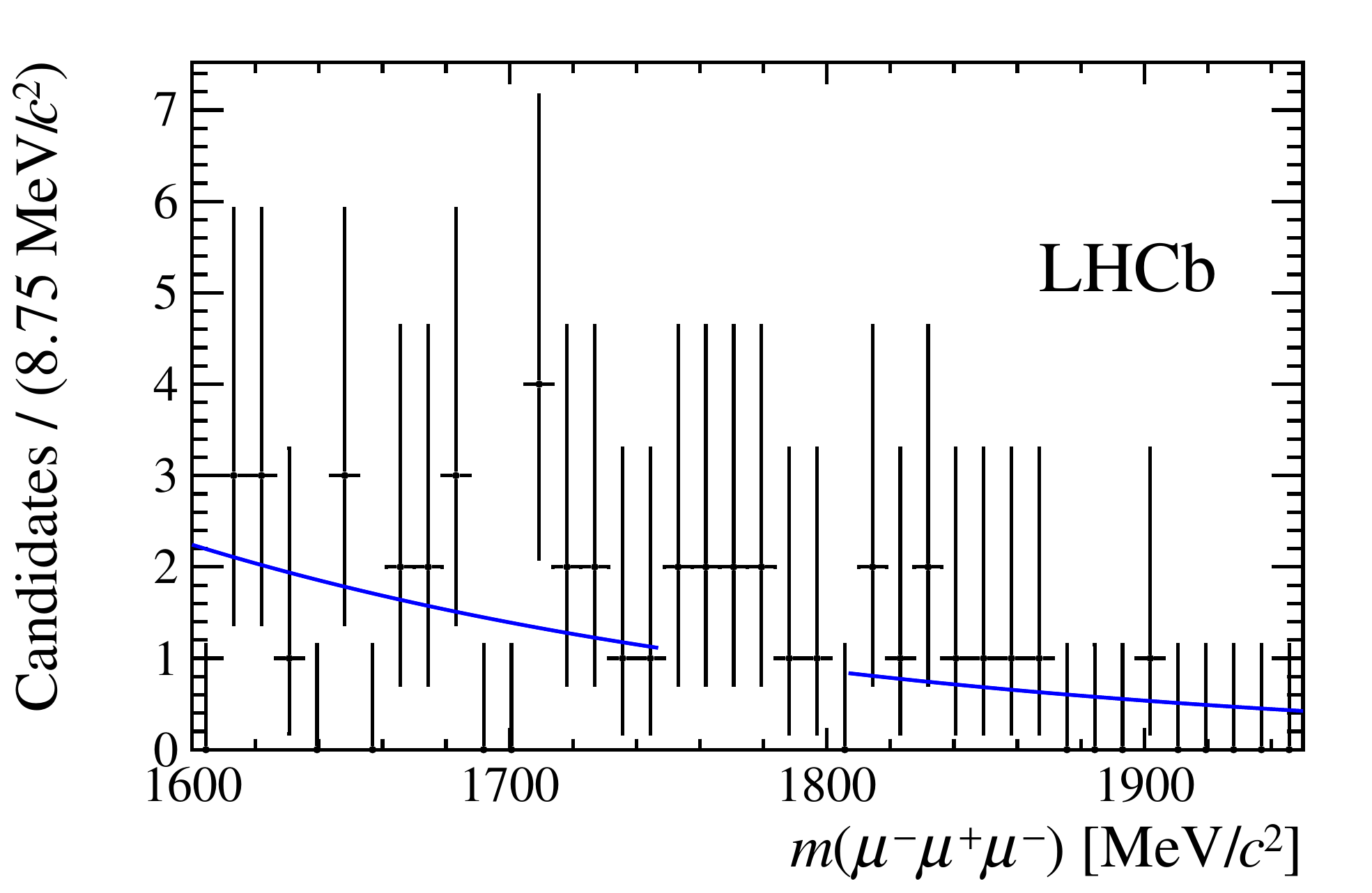}
\caption{Invariant mass distribution for $\Ds\to(\phi\to\mu\mu)\pi^{+}$ (left) and \tmmm (right) candidates. }
\label{fig:tau3mu}
\end{figure}

\section{Conclusions}

The observation of LFV would be a clear sign of new physics and searching for these decays is particularly interesting
because in many models they can be linked to the anomalies observed in LFU tests.
LHCb searched for a variety of such decays: no signal excess is observed and therefore limits are set, which in several cases are the best to date.
As in many channels the level of beyond-SM predictions is within reach, LHCb will continue to analyse these decays including the Run II dataset and
searching for new channels including semileptonic decays such as $\Bp \to K e^{\pm}\mu^{\mp}$ and  $\Lb\to \Lz e^{\pm}\mu^{\mp}$ and decays including $\tau$
leptons such as $\Bz\to\tau^{\pm}\mu^{\mp}$ and $\B\to K^{(*0)}\tau^{\pm}\mu^{\mp}$.

\bibliographystyle{JHEP}
\bibliography{skeleton,LHCB-PAPER,main}

\end{document}